# Artificial Intelligent Ethics in the Digital Era: an Engineering Ethical Framework Proposal

Esteban García-Cuesta
Data Science Laboratory, Faculty of Architecture, Engineering and Design,
Universidad Europea de Madrid, C/Tajo S/N 28670 Villaviciosa de Odón,
Spain

{esteban.garcia@universidadeuropea.es}

## ABSTRACT

Nowadays technology is being adopted on every aspect of our lives and it is one of most important transformation driver in industry. Moreover, many of the systems and digital services that we use daily rely on artificial intelligent technology capable of modeling social or individual behaviors that in turns also modify personal decisions and actions. In this paper, we briefly discuss, from a technological perspective, a number of critical issues including the purpose of promoting trust and ensure social benefit by the proper use of Artificial Intelligent Systems. To achieve this goal we propose a generic ethical technological framework as a first attempt to define a common context towards developing real engineering ethical by design. We hope that this initial proposal to be useful for early adopters and especially for standardization teams.

## EXTENDED ABSTRACT

Artificial Intelligence (AI) as a science has its main goal on the replication of human intelligent behaviors such as reasoning, memorizing, sensing, or solving complex problems. During the last decades, it has shown its success and utility as a general solver at specific applications such as diagnosis, planning, automatic control systems, knowledge search, or prediction (Palma, 2008). Some important abilities that an AI system (AIS) should have in order to achieve that goals are perception, abstraction, recursive reasoning, self-reference, autonomy, and self-decision. The last three are especially important to allow self-improvement and adaptation to environmental changes where the AIS 'lives'.

This is very relevant, and probably one of the big next advances in AI, because it allows the AIS to develop new and unexpected behaviors accessing to thousands of databases and knowledge repositories in Internet and the Web. Examples of those repositories are Wikipedia or DBpedia but also any shared database with a well-defined API for accessing. This data is constantly changing and thereof the AIS should change permanently in order to adapt to those new datasets. AlphaGo (Silver, 2017) is also another example of the capabilities of these systems that are able to develop unexpected strategies as was showed in the AlphaGo challenge.

This implies a big change in current AI algorithms' behavior because we must control (but without doing it explicitly because we then loose some positive capabilities of AIS) how the evolution of a self-learning algorithm occurs assuring the ethical perspective and human values. (Kramer, 2018) indicates the same issue by a set of questions that forces ourselves to make a self-reflection: How we can create and secure the diversity of systems, objectives, and operators?; How to create an overarching framework for the participation-promotion deployment and include of users and state regulatory competence?

## 1. SOCIAL TRANSFORMATION

There are also some other circumstances that should be addressed regarding the ethics on digital transformation and artificial intelligence in particular that has not been commented so far. Search neutrality is a principle that settles that search engines should have no editorial policies other than their results be comprehensive, impartial and based solely on relevance. There are multiple cases that deliberately this search neutrality is violated (i.e. redirection of traffic during searches to show paid announcers) but we want to refer to a more sibylline behavior that appears from the designing of the algorithms itself due to its opacity (Burrel, 2016) and data bias (Bolukbasi, 2016). Nowadays Google's success and its large number of users obliges to pay some attention to its ethical implications. (Zimmer, 2008) says that Google's ultimate goal is to "create 'the perfect search engine' that will provide only intuitive, personalized, and relevant results." Personalized search means the best result suggested by Google's PageRank algorithm (Page, 1999) or its updated and personalized version which may turn out as idiosyncratic and oddly peremptory (Halpern, 2011).

Despite Internet and the Web was initially viewed as a technology that would … 'give voice to diverse social, economic, and cultural groups, to members of society not frequently heard in the public sphere…' (Introna, 2000) it can also describe an 'anti-democratic' aspect of contemporary search technology when is noticed that systematically exclude certain Web sites, or certain type of sites over others pointing out whether these "independent voices and diverse viewpoints" that are essential for a democracy are capable of being "heard through the filter of search engines" (Diaz, 2008). This relates with the description of search engines as "gatekeepers of cyberspace" opening a question about if, having the right to be in the Web is enough to view the Web as a democratic medium? or if, should we expect that the search engines should also assure that the right is accomplished properly by being equally accessible?

## 2. ARTIFICIAL INTELLIGENT SYSTEMS, ETHICS, AND LEGALITY

Extending slightly the definition given at (Poole, 1998) an Artificial Intelligent System (AIS) includes 'any algorithm, or a set of them working together, that perceives its environment and takes actions that maximize its chance of successfully achieving its goals'. Adding the extension of Ethical Artificial Intelligent System (EAIS) we add at the end of that definition '…under some restrictions'.

Then, the next question to be answered is; What are those restrictions or ethical values that an Artificial Intelligent System must have? Following the privacy by design approach to systems engineering (Schaar, 2010; Cavoukian, 2011) we establish the next ethical by design principles in order to maintain an AIS ethical:

1. Preventive not remedial. The ethical protection comes before-the-fact, not after.
2. Ethics embedded into design. The ethics is embedded into the design of the AIS and is an essential component for taking decisions and is integral to the system.
3. Full functionality. Ethical by design must accommodate all possible objectives that an AIS can perform preventing the dichotomy, such as ethical vs. functional, being complementary to all the other functionalities provided by the AIS.
4. Internal and external security. Ethical by design extends securely throughout the entire life of the AIS, strong internal and external security measurements are essential to ethics, from start to finish.
5. Transparency. Ethical by design to assure all stakeholders that they are operating according to the stated defined ethical rules and objectives, and this operation is visible and transparent to users and providers. Trust but verify.
6. Respect for human rights. Ethical by design requires keeping the AIS human centered. It should provide measurements of ethical achievements, behavior notifications, user-friendly, implement self-control and human remote control procedures, and report malfunctioning.

The only way to guarantee that these principles are built on AIS is involving ethical preservation authorities during the whole life of the system since its inception until is deprecation. Auditing studies for algorithms avoiding selective responses should be applied (Sandvig, 2014) and the authorities should also be digital.

In addition, we have identified three technical characteristics that summarize and include all the possibilities and applications that have been defined in European Commission in its Ethics guidelines for trustworthy AI (EC, 2019) plus the above mentioned. The three engineering ethical by design technological characteristics are:

- Transparency: includes all the functionalities that provides traceability, explainability, and accessibility.
- Predictability: includes all the functionalities for verification and testing of actions taken by the AIS at a preventive step for detecting unethical behaviors. Transparency functionalities also assist to predictability.
- Robustness: includes all the functionalities to assure that the ethical values and models defined within the AIS cannot be changed by internal miss behaviors or external attacks that could compromise the whole system, and thereof the ethical module as well.

An AIS that has these characteristics is able to be compliant with all the circumstances related with its regular performance in any context of work and the defined ethical rules. If any characteristic is not present then there is not any warranty regarding the ethical behavior of the AIS. For example, a lack of robustness can be that an external system/person could trespass the security of the AIS and have internal access to the models, data, ethical rules, etc. and hence could modify them (i.e. changing an ethical rule from "don't discriminate anyone by age" to "discriminate anyone by age"). During the study, we have observed a few main ethical concerns in the literature that can be grouped into the next broad categories (that are partially shared with Herman, 2014): (i) search-engine, accessibility bias, and democracy (ii) opacity and algorithmic models bias, and (iii) monitoring, personal privacy and informed consent.

The first points out how relevant is accessing to knowledge and information in an unbiased way. Search personalization has highlighted an alarm because personalized filters act as an invisible autopropaganda indoctrinating us (as individuals and also as societies) with our own ideas and minimizing our exposure to others and living in our own bubble (Pariser, 2011). This has been exemplified by (Datta, 2015) where Google, advertisers, websites, and users context are components of a personalized AI advertisement system. In that case, the gender characteristics of the user will imply a different set of job search answers discriminating by gender current historical data and perpetuating the bias.

The second has its roots on the fact that when a computer learns and consequently builds its own representation of a problem, or model of the world, it does without regard for human comprehension (Burrell, 2016). It could be claimed that auditors who have access to the code can assure its ethical component but this would not be enough because of: the complexity of interpretation of code on execution (the models can change, self-learn and self-adapt) throughout the time, and the limitation of the number of auditors needed to do that work. A combination of transparency and predictability is needed to propose a feasible solution to this issue.

The third has been recently ruled by the General Data Protection Regulation (EU) 2016/679 law and any ethical system must be compliant with it.

## 3. ETHICAL ARTIFICIAL INTELLIGENT SYSTEM FRAMEWORK (EAISF)

To accomplish with the above presented principles and EAIS's characteristics we sketch a first attempt in order to define a general ethical framework in this context.

The framework (Fig.1) has three main parts: (i) the AIS Controller, (ii) the simulation module, and (iii) the ethical evaluation module. **The AIS controller** contains the mechanism that interacts with the outer world. It defines the objectives and establishes a set of well-defined tasks and actions according to the models that it has implemented. This ends up with a set of possible options that the AIS can execute (e.g. turn right or left in AV). **The simulation module** contains a set of components: (i) AIS model that is a replica of the AIS controller that allows to extract the objectives and any other information about the system, (ii) the world model that describes the context where the AIS lives including the set of ethical rules of that world, and the human model that describes the set of ethical rules that defines the human/AIS that is interacting with in that specific context (i.e. an AIS can interact with a human, with another AIS, or with the world). **The ethical evaluation module** verifies that all the specified ethical rules according to the ethical principles are satisfied.

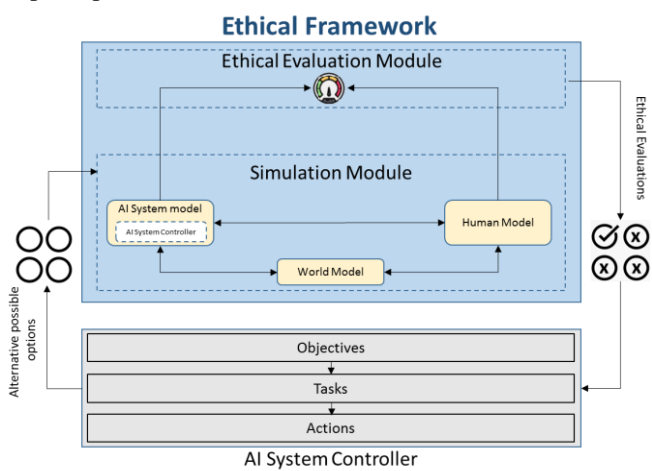

Fig. 1 Description of the proposed Ethical Framework Architecture

The flow of interaction starts (1) at the AIS controller that sends the possible actions/options given its current state to the simulation module. (2) The simulation module receives the different options and has a hierarchical structure that defines more general and domain specific ethical rules including also the person/machine that is interacting with. This permits to adapt its decisions to this information and at the same time to have flexibility in the definition of the different ethical parts. (3) Finally, the informed decision is passed to the ethical evaluation model that verifies the behavior and provides feedback to the AI system controller that executes the action.

Thereof the framework has two different levels for ethical definitions, the general purpose level and domain specific level. This approach is more suitable that a monolithic approach due to most of AIS (we could even say that nowadays all of them) are domain specific (e.g. AIS to help detection of cancer from images and assist the doctor) and can adapt to each one easier without losing the general purpose ethical guides.

In the next, we define technical functionalities that are included in each one of the three proposed characteristics in order to preserve the ethical principles:

- Transparency:
  o the EAIS maintains a historical log of every taken decision or internal change and it should be recorder in a trusted authority.
  o the EAIS provides functionalities to be tested any time by a certified authority.
  o the EAIS provides functionalities to share the models and expected outputs to be stored by trusted and secure authority.
  o the EAIS provides access to ethical reports publicly.
  o the EAIS provides access to reported unethical behaviors.
  o The EAIS provides interfaces for full remote control access by certified authorities.
- Predictability:
  o the EAIS provides functionality to be tested any time for ethical testing purposes by a certified authority without previous notice.
  o the EAIS defines functionalities to test itself regularly in order to verify its correct behavior providing access to the full report.
  o The EAIS reports to a certified authority any failure detected in predictability test.
- Robustness:
  o The EAIS assures its internal security and prevent undesired changes in its models, data, and decision outputs.
  o The EAIS assures its external security and prevent undesired accesses.
  o The EAIS notifies any security fraud to certified authorities.

As can be unveiled, many of the technical parts rely on automatic systems that verify and test them. This is a key part of the proposed general framework because it is not possible to assure that a AIS behaves ethically without using automatic systems that verify it. It is worth highlighting that, as in any system, the security of the whole system is defined by its weakest part. Regarding the certified authorities, the schema relies on the current hierarchical certification authorities' structure and hence the whole system relies in the top certification authority in the hierarchy. Despite we have not talked specifically about security in this paper, we want to point out that it is a masterpiece for the ethical artificial intelligent framework and every technology applicable for security is applicable in the presented framework.

## 4. CONCLUSIONS

The need of a well-defined technological framework to allow the inclusion of ethics in artificial intelligent systems is an important step that has been pointed out by different entities during the last years. Some consequences have been already observed pointing out the relevance of this topic and the social awareness about it. In this paper we have presented, from a technological and engineering points of view, a set of principles that has to be adopted in order to be compliant with the engineering ethical by design concept. We hope that the simplification and summarization of the current EU guides within a few technical characteristics will be useful for early adopters at industry, academia, and standardization teams.

## ACKNOWLEDGMENT

This study has been funded by Universidad Europea de Madrid. We would also like to thank the dean of the Arquitecture, Engineering and Design School Alberto Sols for his helpful discussions and support.